# Robust Tracking Guidance for Zero Propellant Maneuver

Sheng **ZHANG**[1], Qian **ZHAO**[2], Hai-Bing **HUANG**[3],and Guo-Jin **TANG**[4]

**Abstract**:  The Zero Propellant Maneuver (ZPM) maneuvers the space station by large angle, utilizing the Control Momentum Gyroscopes (CMGs) only. A robust tracking guidance strategy is proposed to enhance its performance. It is distinguished from the traditional trajectory tracking guidance in that the reference trajectory is adjusted on-line, under the inspiration of eliminating the discrepancy on the total angular momentum of the space station system. The Lyapunov controller is developed to adjust the attitude trajectory and further redesigned for a better performance based on an interesting physical phenomenon, which is taken advantage of by coupling the components of state vector. The adjusted trajectory is then tracked to reach the target states of maneuver. Simulations results show that the disturbance effects arising from initial state errors, parameter uncertainty and modeling errors are attenuated or even eliminated, which verifies the robustness of the guidance strategy.

[1] Ph.D., Computational Aerodynamics Institution, China Aerodynamics Research and Development Center, Mianyang, People's Republic of China, 621000; zszhangshengzs@hotmail.com.
[2] Ph.D. candidate, College of Aerospace Science and Engineering, National University of Defense Technology, Changsha, People's Republic of China, 410073; zhaoqianmars@gmail.com.
[3] Lecturer, College of Aerospace Science and Engineering, National University of Defense Technology, Changsha, People's Republic of China, 410073; huanghaibing@126.com.
[4] Professor, College of Aerospace Science and Engineering, National University of Defense Technology, Changsha, People's Republic of China, 410073; tang_guojin@hotmail.com.

## 1. Introduction

The Zero Propellant Maneuver (ZPM) becomes widely-known in recent years. This technique realizes large angle attitude maneuvers of space station, using only Control Momentum Gyroscopes (CMGs). Compared with the maneuver realized by the thrusters, the ZPM saves precious propellant and avoids the risk of solar-array contamination. NASA has successfully conducted two ZPM missions on 5 November 2006 and 3 March 2007, when the International Space Station (ISS) was rotated by 90° (Bedrossian et al. 2007a) and 180° (Bedrossian et al. 2007b), respectively. Different from the large angle maneuver executed on other spacecrafts (Ford and Hall 2000; Wie et al. 2002), generally a ZPM transfers the space station from one Torque Equilibrium Attitude (TEA) to another, and the terminal states of attitude, angular velocity and CMGs momentum are all prescribed for momentum management (Bhatt 2007); in particular, in a ZPM the environmental torque is exploited to enable large angle maneuver to be achieved, whilst simultaneously maintaining the CMGs within their operational limit. Since the trajectory planning is crucial to the maneuver, the ZPM is an attitude maneuver guidance problem (Bedrossian et al. 2009).

In principle, the current guidance strategy utilized by ISS is to track the trajectory planned off-line. The executed trajectories of the two ZPM missions were momentum-optimal (Zhang et al. 2014). This type of trajectory possesses the largest CMGs angular momentum redundancy to accommodate the angular momentum deviations arising from various disturbances (Bedrossian et al. 2009). For a ZPM implemented by tracking the trajectory planned off-line, the initial state errors relative to the nominal boundary conditions are inevitable, and they will give discrepancy between the nominal trajectory and the real flight results. Also, the modeling errors will bring deviation of the CMGs momentum profile when tracking the nominal attitude trajectory. These errors will lead to the error in terminal CMGs momentum, which is disadvantageous to the smooth switch to the momentum management, or even worse, lead to the failure of the mission for the lose of CMGs control ability in the case of CMGs momentum saturation. The success of the current ZPM depends heavily on the accuracy of the initial states and the planning model, and there is no guarantee that all are exact when the planned trajectory is actually flown. To ensure the robustness of the maneuver, significant off-line simulations have to be performed to evaluate the impact from various disturbances (Bedrossian et al. 2007a, 2007b). The problem under the current guidance strategy may be solved with the predictive/tracking guidance (Bharadwaj et al. 1998), yet currently the real-time trajectory updating is still hard to achieve on-board.

To enhance the performance of the ZPM, this paper proposes a novel tracking guidance strategy. Different from the traditional trajectory tracking guidance, in the proposed guidance strategy the reference trajectory is adjusted on-line, through a feedback scheme of guiding the total angular momentum of the space station system to the expected value. It may attenuate or even eliminate the disturbance effects arising from initial state errors, parameter uncertainty and modeling errors, which is an important improvement to the current ZPM technique. In the following, Sec. 2 presents the mathematical model. In Sec. 3, the guidance strategy is detailed. In adjusting the attitude trajectory, the Lyapunov controller is developed and further redesigned for a better performance based on an interesting physical phenomenon. The adjusted trajectory is then tracked by a feedback linearization controller. The convergence of the guidance strategy is also discussed. In Sec. 4, ZPM examples are simulated, and the robustness of the guidance strategy is verified.

## 2. Mathematic Model

To derive the equations of motion, relevant reference frames are defined first. The body reference frame, $b$, has its origin at the center of mass of the space station. It is fixed with the space station and its axes are aligned with the geometric characteristic directions. The Local Vertical Local Horizontal (LVLH) orbit reference frame, $o$, has origin $o_o$ that coincides with the center of mass of the space station. The $o_o z_o$ axis is aligned with the local vertical, towards the center of Earth, the $o_o x_o$ axis lies on the orbit plane in the transverse direction, normal to $o_o z_o$, and the $o_o y_o$ axis is perpendicular to the orbit plane, completing a right-handed triad. The orbit frame makes one rotation about the Earth during each orbit period. In the paper, a circular orbit is assumed for the space station, so that the orbit rotation rate, $n$, is constant. The inertial frame, $i$, coincides with the initial orbit frame, but is fixed in the inertial space. For the same quantity described in different frames, a superscript is used to denote the specific frame, and it is omitted when the quantity is described in the body frame.

The Modified Rodrigues Parameters (MRPs) are the minimal description of attitude, which avoids singularities for a principal rotation up to $\pm 360$ deg (Schaub and Junkins 2003). They are defined as

$$\boldsymbol{\sigma} = \begin{bmatrix} \sigma_1 & \sigma_2 & \sigma_3 \end{bmatrix}^T = \boldsymbol{e} \tan \frac{\theta}{4} \qquad (1)$$

where $\boldsymbol{e}$ is the principal rotation axis and $\theta$ is the principal rotation angle. The kinematic equation which describes the attitude of the space station with respect to the orbit is

$$\dot{\boldsymbol{\sigma}} = \boldsymbol{T}(\boldsymbol{\sigma})\big(\boldsymbol{\omega} - \boldsymbol{\omega}_o\big) \tag{2}$$

where $\boldsymbol{T}(\boldsymbol{\sigma})$ is the kinematic matrix, $\boldsymbol{\omega}$ and $\boldsymbol{\omega}_o = \boldsymbol{R}_o^b(\boldsymbol{\sigma})\begin{bmatrix} 0 & -n & 0 \end{bmatrix}^T$ are the space station angular velocity and the orbit frame angular velocity, described in the body frame, respectively, $\boldsymbol{R}_o^b$ is the rotation matrix from the orbit frame, $o$, to the body frame, $b$. The specific form of $\boldsymbol{T}(\boldsymbol{\sigma})$ and $\boldsymbol{R}_o^b$ are given in Schaub and Junkins (2003).

The dynamic equation described in the body reference frame is

$$\dot{\boldsymbol{\omega}} = \boldsymbol{J}^{-1}\big(\boldsymbol{\tau}_e - \boldsymbol{u} - \boldsymbol{\omega} \times (\boldsymbol{J}\boldsymbol{\omega})\big) \tag{3}$$

where $\boldsymbol{J}$ is the inertia matrix of the space station, $\boldsymbol{u}$ is the control generated by the CMGs, $\boldsymbol{\tau}_e$ is the environmental torque, and the "$\times$" denotes the vector cross product.

The motion of the CMGs must also be considered in a ZPM, because of their limited capacity and the boundary condition constraints. The equation of motion of the CMGs is

$$\dot{\boldsymbol{h}}_c = \boldsymbol{u} - \boldsymbol{\omega} \times \boldsymbol{h}_c \tag{4}$$

where $\boldsymbol{h}_c$ is the angular momentum of the CMGs described in the body frame. CMGs have limits on their angular momentum and torque. Hence, during a maneuver the CMGs must operate within their performance range, which may be written as constraints on the angular momentum and the rate of angular momentum change (Bhatt 2007) as

$$\left\| \boldsymbol{h}_c \right\| \le h_{\max} \tag{5}$$

and

$$\left\| \dot{\boldsymbol{h}}_c \right\| \le \dot{h}_{\max} \tag{6}$$

for threshold parameters $h_{\max}$ and $\dot{h}_{\max}$.

The total angular momentum of the space station system is

$$\boldsymbol{H} = \boldsymbol{h}_c + \boldsymbol{h}_s \tag{7}$$

where $\boldsymbol{h}_s = \boldsymbol{J}\boldsymbol{\omega}$. Being observed in the inertial space, $\boldsymbol{H}^i$ may change greatly during the maneuver. Because the CMGs are momentum exchange devices, they cannot generate variation of the total angular momentum. Thus, the environmental torque must be exploited to realize a ZPM. The differential equation of the total angular momentum, described in the orbit frame, is

$$\dot{\boldsymbol{H}}^o = -[\boldsymbol{\omega}_o^o \times]\boldsymbol{H}^o + \boldsymbol{\tau}_e^o \tag{8}$$

where $\boldsymbol{\omega}_o^o = \begin{bmatrix} 0 & -n & 0 \end{bmatrix}^T$ is the orbit frame angular velocity described in the orbit frame. $[\boldsymbol{\omega}_o^o \times]$ is the skew symmetric matrix defined as $[\boldsymbol{\omega}_o^o \times]\boldsymbol{H}^o = \boldsymbol{\omega}_o^o \times \boldsymbol{H}^o$ .

## 3. Robust Tracking Guidance strategy

Before the robust tracking guidance strategy is presented, the traditional tracking guidance is considered first to show how the CMGs momentum error arises. For the trajectory planned off-line, the initial state errors at the beginning of maneuver are inevitable. In Bhatt (2007), for the CMGs initial momentum error, based on the simulation results, it is found that the relationship between the initial CMGs momentum error magnitude and the peak momentum is close to linear, and the final CMGs momentum error is also well-correlated with the initial momentum error. This phenomenon may be explained theoretically for the traditional tracking guidance. For the following explanation, introduce the momentum increment, $\Delta \boldsymbol{H}^i = \boldsymbol{H}^i - \boldsymbol{H}_0^i$ , where $\boldsymbol{H}_0^i$ is the total angular momentum at the maneuver startup time point $t_0$ . $\Delta \boldsymbol{H}^i$ describes the increase of angular momentum in the inertial space, and it satisfies

$$\Delta \dot{\boldsymbol{H}}^i = \boldsymbol{\tau}_e^i(\boldsymbol{\sigma}, t) \tag{9}$$

For $\Delta \boldsymbol{H}^i$ , its initial value is zero and its profile is determined by the environmental torque, which is dominated by the attitude trajectory. The real maneuver conforms to the momentum relation

$$\boldsymbol{h}_c^i(t) + \boldsymbol{h}_s^i(t) = \boldsymbol{h}_c^i(t_0) + \boldsymbol{h}_s^i(t_0) + \Delta \boldsymbol{H}^i(t) \tag{10}$$

Using the tilde '$\sim$' to denote the results planned off-line under nominal conditions, there also is

$$\tilde{\boldsymbol{h}}_c^i(t) + \tilde{\boldsymbol{h}}_s^i(t) = \tilde{\boldsymbol{h}}_c^i(t_0) + \tilde{\boldsymbol{h}}_s^i(t_0) + \Delta \tilde{\boldsymbol{H}}^i(t) \tag{11}$$

If the initial attitudes and angular velocities of the nominal trajectory and the real maneuver are same, then $\boldsymbol{h}_s^i(t_0) = \tilde{\boldsymbol{h}}_s^i(t_0)$ . Presuming that the modeling is accurate and the nominal attitude trajectory is tracked precisely, then $\boldsymbol{h}_s^i(t) = \tilde{\boldsymbol{h}}_s^i(t)$ and $\Delta \boldsymbol{H}^i(t) = \Delta \tilde{\boldsymbol{H}}^i(t)$ . Subtracting Eq. (11) from (10) and denoting the CMGs angular momentum error as $\boldsymbol{\delta}_{h_c} = \boldsymbol{h}_c - \tilde{\boldsymbol{h}}_c$ gives

$$\boldsymbol{\delta}_{h_c}^i(t) = \boldsymbol{h}_c^i(t) - \tilde{\boldsymbol{h}}_c^i(t) = \boldsymbol{h}_c^i(t_0) - \tilde{\boldsymbol{h}}_c^i(t_0) = \boldsymbol{\delta}_{h_c}^i(t_0) \tag{12}$$

Eq. (12) shows that the CMGs momentum error $\boldsymbol{\delta}_{h_c}$ is not changed but maintains $\boldsymbol{\delta}^i{}_{h_c}(t_0)$ in the inertial space during the maneuver. Thus its magnitude is constant. When the CMGs peak angular momentum vector is in the same direction to $\boldsymbol{\delta}_{h_c}$, the relation between the initial momentum error magnitude and the peak momentum is linear. As pointed out in Bhatt (2007): the peak momentum seems to be just the initial CMGs momentum error added to the nominal peak momentum. Eq. (12) also explains why the final CMGs momentum error is linearly-correlated with the initial momentum error. For errors in the initial attitude and angular velocity, they may be transformed to the error of CMGs momentum during the tracking process. Besides the errors of initial states, the modeling error will also result in the deviation of CMGs momentum profile in the maneuver. According to Eq. (7), under precise tracking of the attitude trajectory, the error of the CMGs momentum relative to the planned profile, $\boldsymbol{\delta}_{h_c}$, equals the error of the total momentum, $\boldsymbol{\delta}_{H^o} = \boldsymbol{H}^o - \tilde{\boldsymbol{H}}^o$, i.e.,

$$\boldsymbol{\delta}_{h_c} = \boldsymbol{R}_o^b \boldsymbol{\delta}_{H^o} \tag{13}$$

Now the principle of the robust tracking guidance is introduced. Consider the equations of motion given by Eqs. (2), (3), and (8). They form a cascaded model. For a ZPM with fixed terminal states, that the total momentum of the space station system, $\boldsymbol{H}$, reaches the prescribed value is the necessary condition for the completion of ZPM. Viewing the ZPM as guiding the total momentum to the expected state gives insight to the principle of ZPM. From Eq. (8), it is shown that the change of the total momentum is determined by the environmental torque, which is regulated by the attitude profile. Thus, according to Eq. (13), if the reference attitude trajectory could be adjusted in a feedback manner according to the error of total momentum during the tracking process, then the deviation of CMGs momentum may be gradually reduced or even eliminated. Motivated by the idea, the robust guidance strategy is proposed and the block diagram is presented in Fig. 1. The main difference from the traditional tracking guidance is that a trajectory adjusting controller is set, which determines the adjustment quantities of the attitude trajectory. These quantities are added to the nominal trajectory to obtain the adjusted attitude trajectory. The trajectory tracking controller tracks the trajectory and generates the control command to the CMGs.

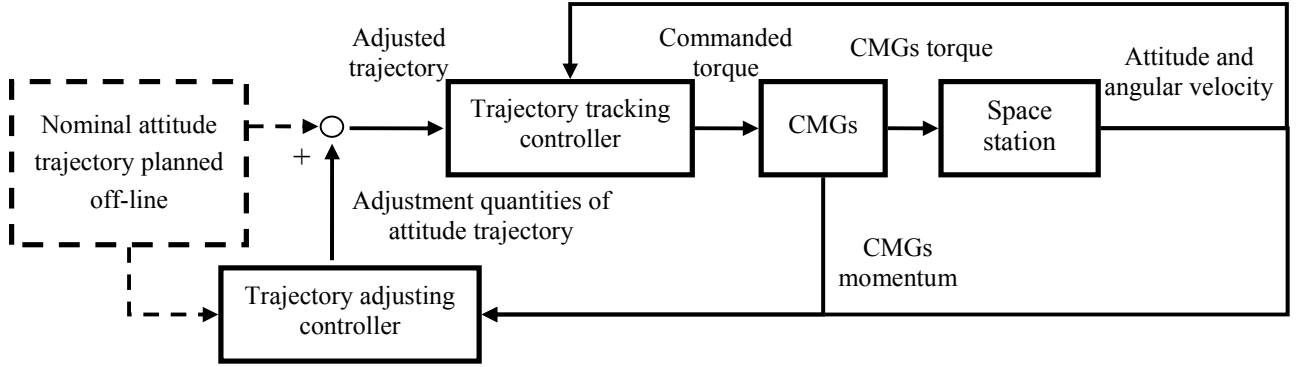

**Fig. 1. Block diagram of the robust tracking guidance**

### 3.1 Trajectory Adjusting Controller

Since the earth gravity gradient torque, $\boldsymbol{\tau}_{gg}$, and the aerodynamic torque, $\boldsymbol{\tau}_a$, are the primary environmental torques acting on the space station, other environmental torques are regarded as disturbance during the controller design. Consider the total momentum differential equation given by Eq. (8). The linear time-varying form may be obtained through expanding it along the nominal trajectory

$$\dot{\boldsymbol{\delta}}_{H^o} = -[\boldsymbol{\omega}_o^o \times]\boldsymbol{\delta}_{H^o} + \boldsymbol{C}\boldsymbol{\delta}_\sigma + \boldsymbol{d} \tag{14}$$

where $\boldsymbol{\delta}_\sigma$ is the adjustment attitude, regarded as the control in Eq. (14), $\boldsymbol{C} = (\boldsymbol{\tau}_{gg}^o)_\sigma + (\boldsymbol{\tau}_a^o)_\sigma$ is the time-varying matrix calculated based on the nominal trajectory, $\boldsymbol{d}$ is the disturbance including other disturbance torques and the higher order expansion terms of the gravity gradient torque and the aerodynamic torque.

For the nominal system without disturbance

$$\dot{\boldsymbol{\delta}}_{H^o} = -[\boldsymbol{\omega}_o^o \times]\boldsymbol{\delta}_{H^o} + \boldsymbol{C}\boldsymbol{\delta}_\sigma \tag{15}$$

Construct the Lyapunov function

$$V = \frac{1}{2}(\boldsymbol{\delta}_{H^o})^T \boldsymbol{\delta}_{H^o} \tag{16}$$

To guarantee its derivative non-positive, i.e., $\dot{V} \le 0$, the feedback control law can be selected as

$$\boldsymbol{\delta}_\sigma = -\boldsymbol{K}_a \boldsymbol{C}^T \boldsymbol{\delta}_{H^o} \tag{17}$$

where $\boldsymbol{K}_a = k_a \boldsymbol{I}$ is the gain matrix, $k_a$ is a positive constant and $\boldsymbol{I}$ is 3×3 identity matrix. Considering the frequency response of the space station and avoiding excessive attitude adjustment, $k_a$ is set to be small. Especially, an interesting phenomenon is found that when the orbit rate, $n$, changes, the decrease of $V$ may be speeded at the end of a finite time horizon. Since $n$ is a constant, an alternative way to approximate the effect is to couple the $x$ and $z$ components of $\boldsymbol{\delta}_{H^o}$, by introducing the following control form

$$\boldsymbol{\delta}_\sigma = -\boldsymbol{K}_a \boldsymbol{C}^T \boldsymbol{R} \boldsymbol{\delta}_{H^o} \tag{18}$$

where $\boldsymbol{R} = \begin{bmatrix} 1 & 0 & -k_{r1} \\ 0 & 1 & 0 \\ k_{r2} & 0 & 1 \end{bmatrix}$, $k_{r1}$ and $k_{r2}$ are positive constants. Since the coupling may be harmful to the contraction of the phase flow of $\boldsymbol{\delta}_{H^o}$, there is a compromise when setting $k_{r1}$ and $k_{r2}$. Note that when $k_{r1} = k_{r2}$, the contraction of the phase flow is not influenced according to the Liouville theorem (Arnold 1992). Currently the matrix $\boldsymbol{R}$ is obtained by trial and error.

Now consider the impact from the disturbance $\boldsymbol{d}$. Using the controller given by Eq. (17), it may be derived that the system described by Eq. (14) is boundedly stable through the Lyapunov analysis. When employing the controller of Eq (18), the Lyapunov method is not applicable. However, according to the liner system theory (Cheng 1999), the faster decrease of $\boldsymbol{\delta}_{H^o}$ is advantageous to repelling the disturbance $\boldsymbol{d}$, because the state transition matrix may attenuate the effect of the disturbance more effectively. The controllers given by Eqs. (17) and (18) are called the Lyapunov Trajectory Adjusting Controller (LTAC) and the Redesigned Trajectory Adjusting Controller (RTAC), respectively. Obviously, LTAC is a special case of RTAC when $k_{r1}$ and $k_{r2}$ equal zero.

Besides $\boldsymbol{\delta}_\sigma$, its first and second order derivatives are also needed for tracking. Differentiating Eq. (18) gives

$$\dot{\boldsymbol{\delta}}_\sigma = -\boldsymbol{K}_a (\boldsymbol{B}_1 \boldsymbol{\delta}_{H^o} + \boldsymbol{B}_2 \boldsymbol{\delta}_\sigma) \tag{19}$$

$$\ddot{\boldsymbol{\delta}}_\sigma = -\boldsymbol{K}_a (\boldsymbol{B}_3 \boldsymbol{\delta}_{H^o} + \boldsymbol{B}_4 \boldsymbol{\delta}_\sigma + \boldsymbol{B}_2 \dot{\boldsymbol{\delta}}_\sigma) \tag{20}$$

where $\boldsymbol{B}_1 = \left(\dfrac{\mathrm{d}\boldsymbol{C}}{\mathrm{d}t}\right)^T \boldsymbol{R} - \boldsymbol{C}^T \boldsymbol{R}[\boldsymbol{\omega}_o^o \times]$, $\boldsymbol{B}_2 = \boldsymbol{C}^T \boldsymbol{R} \boldsymbol{C}$, $\boldsymbol{B}_3 = \left(\dfrac{\mathrm{d}^2\boldsymbol{C}}{\mathrm{d}t^2}\right)^T \boldsymbol{R} - 2\left(\dfrac{\mathrm{d}\boldsymbol{C}}{\mathrm{d}t}\right)^T \boldsymbol{R}[\boldsymbol{\omega}_o^o \times] + \boldsymbol{C}^T \boldsymbol{R}[\boldsymbol{\omega}_o^o \times]^2$,

$\boldsymbol{B}_4 = 2\left(\dfrac{\mathrm{d}\boldsymbol{C}}{\mathrm{d}t}\right)^T \boldsymbol{R} \boldsymbol{C} - \boldsymbol{C}^T \boldsymbol{R}[\boldsymbol{\omega}_o^o \times]\boldsymbol{C} + \boldsymbol{C}^T \boldsymbol{R}\left(\dfrac{\mathrm{d}\boldsymbol{C}}{\mathrm{d}t}\right)^T$, $\dfrac{\mathrm{d}\boldsymbol{C}}{\mathrm{d}t}$, $\dfrac{\mathrm{d}^2\boldsymbol{C}}{\mathrm{d}t^2}$ are the coefficient matrixes computed base on the nominal attitude trajectory. Thus, the adjusted trajectory (denoted by '^') is

$$\hat{\sigma} = \tilde{\sigma} + \boldsymbol{\delta}_\sigma \tag{21}$$

$$\dot{\hat{\sigma}} = \dot{\tilde{\sigma}} + \dot{\boldsymbol{\delta}}_\sigma \tag{22}$$

$$\ddot{\hat{\sigma}} = \ddot{\tilde{\sigma}} + \ddot{\boldsymbol{\delta}}_\sigma \tag{23}$$

where $\tilde{\sigma}$, $\dot{\tilde{\sigma}}$ and $\ddot{\tilde{\sigma}}$ are the nominal results planned off-line.

### 3.2 Trajectory Tracking Controller

The trajectory tracking controller tracks the adjusted trajectory. To design the tracking controller, the feedback linearization technique (not limited to) is utilized. It transforms the nonlinear system into a normal canonical form that behaves linearly without approximation (Khalil 2002). Consider the kinematic and dynamic equations given by Eqs. (2) and (3), respectively. Regarding the MRPs as the output functions, the system could be feedback linearized as

$$\frac{\mathrm{d}}{\mathrm{d}t}\begin{bmatrix} \sigma_1 \\ \dot{\sigma}_1 \\ \sigma_2 \\ \dot{\sigma}_2 \\ \sigma_3 \\ \dot{\sigma}_3 \end{bmatrix} = \begin{bmatrix} 0 & 1 & & & & \\ 0 & 0 & & & & \\ & & 0 & 1 & & \\ & & 0 & 0 & & \\ & & & & 1 & 0 \\ & & & & 0 & 0 \end{bmatrix}\begin{bmatrix} \sigma_1 \\ \dot{\sigma}_1 \\ \sigma_2 \\ \dot{\sigma}_2 \\ \sigma_3 \\ \dot{\sigma}_3 \end{bmatrix} + \begin{bmatrix} 0 \\ 1 \\ 0 \\ 1 \\ 0 \\ 1 \end{bmatrix}\boldsymbol{v} \tag{24}$$

where

$$\begin{aligned} \boldsymbol{v} &= \ddot{\sigma} \\ &= \frac{\mathrm{d}\boldsymbol{T}(\sigma)}{\mathrm{d}t}(\boldsymbol{\omega} - \boldsymbol{\omega}_o) + \boldsymbol{T}(\sigma)\left( \boldsymbol{J}^{-1}\left( \boldsymbol{\tau}_e - \boldsymbol{\omega}\times(\boldsymbol{J}\boldsymbol{\omega}) - \boldsymbol{u}\right) - \frac{\mathrm{d}\boldsymbol{R}_o^b}{\mathrm{d}t}\boldsymbol{\omega}_o^o \right) \end{aligned} \tag{25}$$

is the transformed control.

The tracking error to the adjusted trajectory is $\Delta\boldsymbol{\sigma} = \boldsymbol{\sigma} - \hat{\boldsymbol{\sigma}}$ and $\Delta\dot{\boldsymbol{\sigma}} = \dot{\boldsymbol{\sigma}} - \dot{\hat{\boldsymbol{\sigma}}}$. The three independent subsystems included in Eq. (24) could be considered separately. Taking the first subsystem for example, the error dynamic equation is

$$\frac{\mathrm{d}}{\mathrm{d}t}\begin{bmatrix} \Delta\sigma_1 \\ \Delta\dot{\sigma}_1 \end{bmatrix} = \begin{bmatrix} 0 & 1 \\ 0 & 0 \end{bmatrix}\begin{bmatrix} \Delta\sigma_1 \\ \Delta\dot{\sigma}_1 \end{bmatrix} + \begin{bmatrix} 0 \\ 1 \end{bmatrix}\Delta v_1 \tag{26}$$

where $\Delta v_1$ is the first component of $\Delta\boldsymbol{v} = \ddot{\boldsymbol{\sigma}} - \ddot{\hat{\boldsymbol{\sigma}}}$. The feedback controller takes the form

$$\Delta v_1 = k_p^1 \Delta\sigma_1 + k_d^1 \Delta\dot{\sigma}_1 \tag{27}$$

where $k_p^1$ and $k_d^1$ are the gain coefficients for the first subsystem. Achieving $\Delta\sigma$, $\Delta\dot{\sigma} \to 0$ requires the poles of the feedback system being located in the open left-half plane. Also, the flexible frequency of the space station and the response requirement needs to be considered when setting the gains.

Feedback linearization has a great disadvantage that the exactness of the model is required (Sheen and Bishop 1994). The controller given by Eq. (27) may be augmented to be a PID controller to address the error in modeling. However, since the momentum of the CMGs is also expected to be tracked, the integral term, which may sacrifice the accuracy of the CMGs momentum, is not integrated during the maneuver. Other means such as the adaptive control technique (Schaub et al. 2001) or the sliding mode control technique (Cong et al. 2014) may be employed to address the deficiency of feedback linearization. This will be studied in the future and is not considered here. For the whole system given by Eq. (24), the feedback controller is

$$\Delta\boldsymbol{v} = \boldsymbol{K}_t \begin{bmatrix} \Delta\sigma_1 & \Delta\dot{\sigma}_1 & \Delta\sigma_2 & \Delta\dot{\sigma}_2 & \Delta\sigma_3 & \Delta\dot{\sigma}_3 \end{bmatrix}^T \tag{28}$$

where $\boldsymbol{K}_t = \begin{bmatrix} k_p^1 & k_d^1 & & & & \\ & & k_p^2 & k_d^2 & & \\ & & & & k_p^3 & k_d^3 \end{bmatrix}$ is the gain matrix.

The control for trajectory tracking includes the feedforward control, $\bar{\boldsymbol{v}} = \ddot{\bar{\boldsymbol{\sigma}}}$, and the feedback control, i.e.,

$$\boldsymbol{v} = \bar{\boldsymbol{v}} + \Delta\boldsymbol{v} \tag{29}$$

According the inverse control transformation, the CMGs commanded torque is

$$\boldsymbol{u}_{cmd} = \boldsymbol{\tau}_e - \boldsymbol{\omega} \times (\boldsymbol{J}\boldsymbol{\omega}) - \boldsymbol{J}\boldsymbol{T}(\boldsymbol{\sigma})^{-1} \left( -\frac{\mathrm{d}\boldsymbol{T}(\boldsymbol{\sigma})}{\mathrm{d}t}(\boldsymbol{\omega} - \boldsymbol{\omega}_o) + \boldsymbol{T}(\boldsymbol{\sigma})\frac{\mathrm{d}\boldsymbol{R}_o^b}{\mathrm{d}t}\boldsymbol{\omega}_o^o + \boldsymbol{v} \right) \tag{30}$$

Since $\boldsymbol{T}(\boldsymbol{\sigma})^{-1} = \dfrac{16}{(1 + \boldsymbol{\sigma}^T\boldsymbol{\sigma})^2}\boldsymbol{T}^T(\boldsymbol{\sigma})$ (Schaub and Junkins 2003), the transformation is guaranteed within the normal range of MRPs.

### 3.3 Convergence of Guidance Strategy

For the robust tracking guidance strategy, error in the total angular momentum will be reduced or even eliminated under the trajectory adjusting controller, and the convergence to the nominal trajectory is guaranteed by the following proposition.

**Proposition 1:** For the guidance strategy given by Fig.1, if the total angular momentum converges to the nominal profile, then the attitude, angular velocity and the momentum of the CMGs will converge to the nominal results.

**Proof:** If the total angular momentum converges to the nominal profile, then the adjustment quantities of trajectory tend to zero. Thus, the attitude and angular velocity will converge to the nominal results under the control of tracking controller. Also, according to the momentum relation given by Eq. (7), the momentum of the CMGs will converge to the nominal result.

$$\square$$

The path constraints originating from the CMGs performance limitation are not specialized in the guidance strategy proposed. Thus crucial questions are now raised. Will the CMGs momentum be saturated under various disturbances during the maneuver? What will happen if the CMGs momentum is saturated? It is possible that the CMGs momentum saturation occurs under various disturbances. However, the essence of the robust ZPM guidance is to reduce the discrepancy of the total angular momentum profile. As long as the error in the total angular momentum vanishes, Proposition 1 guarantees that the CMGs momentum will converges to the nominal, thus the CMGs momentum saturation will be gotten rid of ultimately.

## 4 Simulation Examples

The $-90$ deg ZPM mission taken from Bhatt (2007) will be used. The orbital rotation rate is $n = 1.1461 \times 10^{-3}$ rad/s, and the inertia matrix of the space station is

$$\boldsymbol{J} = \begin{bmatrix} 24180443 & 3780010 & 3896127 \\ 3780010 & 37607882 & -1171169 \\ 3896127 & -1171169 & 51562389 \end{bmatrix} \text{ kg m}^2 \tag{31}$$

The constraints for the CMGs are a maximum momentum of $h_{\max} = 19524$ Nms and a maximum rate of change of momentum of $\dot{h}_{\max} = 271.16$ Nm. A simplified aerodynamic model is utilized. The aerodynamic drag was computed based on the projected cross-sectional area of the spacecraft in the direction of the relative wind of a rotating atmosphere. The cross-sectional area is assumed to be a constant of $500$ m$^2$. The vector from the mass center to the pressure center is assumed to be fixed in the body frame, and given by $[-9.70, 1.71, 1.74]^T$ m. The mass density of the atmosphere is $2 \times 10^{-11}$ kg/m$^3$, and the drag coefficient is 2.2. The initial time is $t_0 = 0$ s and the terminal time is $t_f = 6000$ s. Table 1 gives the nominal initial and terminal boundary conditions.

**Table 1. Nominal initial and terminal boundary conditions for the ZPM mission**

| Initial state | Value | Terminal state | Value |
|---|---|---|---|
| $\boldsymbol{\sigma}(t_0)$ | $[0.1352, -0.4144, 0.5742]^T \times 10^{-1}$ | $\boldsymbol{\sigma}(t_f)$ | $[-0.3636, -0.2063, -4.1360]^T \times 10^{-1}$ |
| $\boldsymbol{\omega}(t_0)$ (rad/s) | $[-0.2541, -1.1145, 0.0826]^T \times 10^{-3}$ | $\boldsymbol{\omega}(t_f)$ (rad/s) | $[1.1353, 0.0030, -0.1571]^T \times 10^{-3}$ |
| $\boldsymbol{h}_c(t_0)$ (Nms) | $[-0.6725, -0.2373, -5.2768]^T \times 10^3$ | $\boldsymbol{h}_c(t_f)$ (Nms) | $[-0.0122, -4.8226, -0.1830]^T \times 10^3$ |

The nominal trajectory planned off-line is the momentum-optimal type using GPOPS (Patterson and Rao 2015). Since the momentum-optimal trajectory has a large rate of change of the CMGs momentum around the initial and terminal times, in the nominal trajectory planning, the rate of momentum change constraint of the CMGs is strengthened for a conservative performance, with a smaller threshold parameter of $0.8\,\dot{h}_{\max}$.

The simulation integrator is ode45 with a relative error tolerance of $1 \times 10^{-6}$ under Matlab. For the double gimbal CMGs, since the gradient steering law reacts well to answer the commanded torque (Yoshikawa 1975), the steering law is not considered in the simulation. However, the angular momentum constraint and the rate of momentum change constraint are considered, and when the momentum of CMGs is saturated, only the control towards the inside of momentum envelope is allowed. For the trajectory adjusting controller, $k_a = 5 \times 10^{-8}$, and $k_{r1} = k_{r2} = 1.6$. For the trajectory tracking controller, referring to the setting of the PD attitude hold controller on ISS utilized in Bhatt (2007), the control bandwidth $\omega_n$ and the damping ratio $\zeta$ of the feedback system are set to be 0.01 and 0.707, respectively. The gain coefficients of $\boldsymbol{K}_t$ of the tracking controller are then calculated by

$$k_p^i = -\omega_n^2, k_d^i = -2\omega_n\zeta, i = 1, 2, 3 \tag{32}$$

### 4.1 Maneuver under initial state errors

In this subsection, maneuvers with initial state errors are investigated. First consider a ZPM, in which the initial attitude and angular velocity equal the nominal values exactly, while the initial error for the momentum of the CMGs is $\boldsymbol{\delta}_{h_c}(t_0) = [4500, 0, 0]^T$ Nms. This may be achieved by applying the attitude hold controller to eliminate the errors in attitude and angular velocity at the maneuver startup. Table 2 presents the terminal state errors of simulations. The error of attitude is represented by the principal rotation angle. It is shown that the robust tracking guidance especially with RTAC performs well to eliminate the error of the CMGs momentum.

**Table 2. Terminal state errors of ZPM under initial CMGs momentum error of [4500,0,0]$^\text{T}$ Nms**

| Tracking Strategy | Principal rotation angle of attitude error (deg) | Magnitude of angular velocity error (rad/s) | Magnitude of CMGs momentum error (Nms) | Magnitude of total momentum error (Nms) |
|---|---|---|---|---|
| Tradition tracking | $5.2639 \times 10^{-4}$ | $3.8473 \times 10^{-8}$ | 4500 | 4500 |
| Robust tracking with LTAC | 0.1156 | $2.4888 \times 10^{-6}$ | $2.6824 \times 10^{2}$ | $2.0101 \times 10^{2}$ |
| Robust tracking with RTAC | 0.0023 | $4.1276 \times 10^{-8}$ | 0.6874 | 1.1251 |

Figures 2-4 compare the attitude results. It is shown that, for the traditional tracking guidance the simulation results coincide with the nominal trajectory, while for the robust tracking guidance the attitude trajectory differs but gradually converges to the nominal trajectory. By adjusting the attitude trajectory on-line, the error of CMGs momentum is eliminated with the robust tracking guidance strategy.

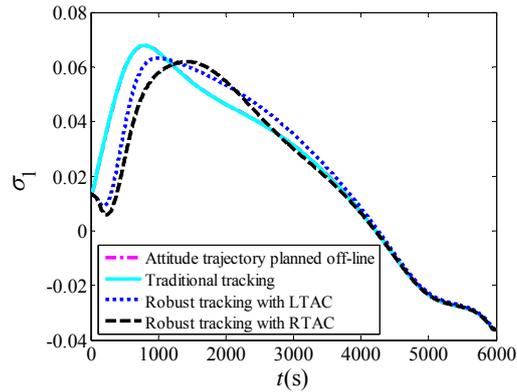

**Fig. 2. Simulation results for $\sigma_1$ under the initial CMGs momentum error**

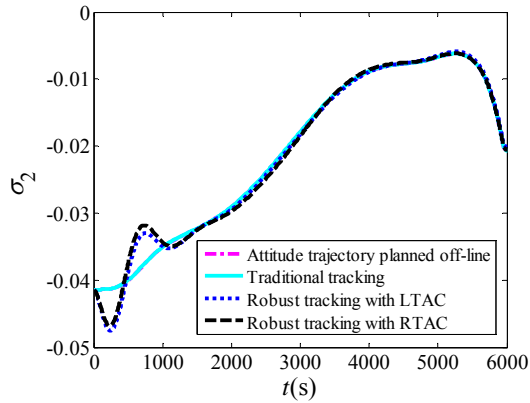

**Fig. 3. Simulation results for $\sigma_2$ under the initial CMGs momentum error**

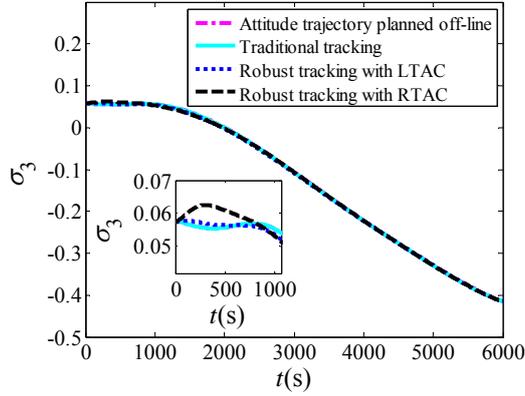

**Fig. 4. Simulation results for $\sigma_3$ under the initial CMGs momentum error**

Figure 5 presents the momentum magnitude of the CMGs. It is shown that the error holds for the traditional tracking, while it gradually decreases for the robust tracking guidance.

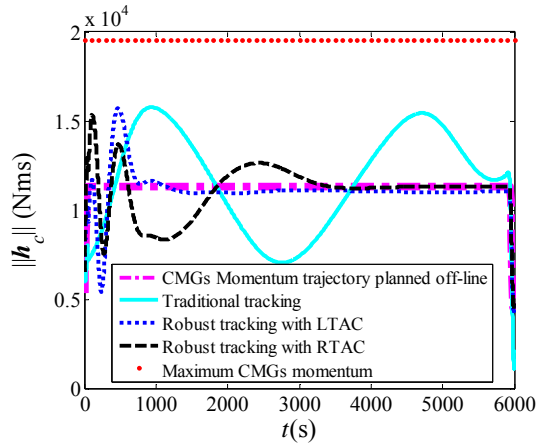

**Fig. 5. Simulation results of the CMGs momentum magnitude under the initial CMGs momentum error**

The relation curve between the magnitude of total momentum error and the maneuver time is presented in Fig. 6. The error is constant for the traditional tracking, which means the magnitude of CMGs momentum error is constant during the maneuver. For the robust tracking guidance with LTAC, $\left\| \boldsymbol{\delta}_{H''} \right\|$ monotonically decreases as the space station maneuvers. For the tracking guidance with RTAC, although the error temporarily increases in the middle, the performance is better.

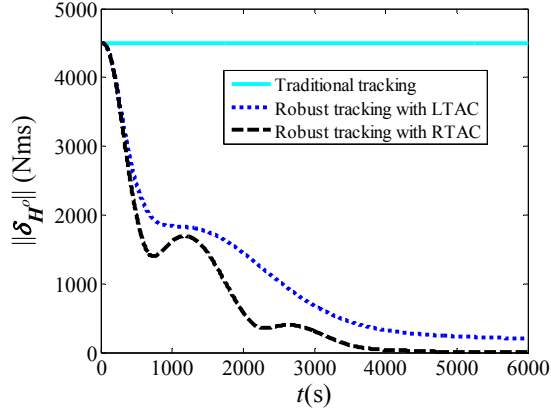

**Fig. 6. Simulation results of the total momentum error magnitude under the initial CMGs momentum error**

To further investigate the performance of the guidance strategies under various errors of initial states, Monte-Carlo simulations with 100 samples were performed. The error of attitude is expressed by the principal rotation axis and the principal rotation angle; the direction of the principal rotation axis is random and the principal rotation angle is 5 deg. For the error of angular velocity, the direction is random and the magnitude is 0.05 $n$, where $n$ is the orbit rotation rate. For the error of CMGs momentum, the direction is random and the magnitude is 1000 Nms. Table 3 presents the terminal state errors. It shows that the traditional tracking has a large CMGs momentum error, while the robust tracking guidance especially with RTAC can achieve the maneuver target effectively.

**Table 3. Terminal state errors of ZPM under initial state errors**

| Tracking Strategy | Principal rotation angle of attitude error (deg) | | Magnitude of angular velocity error (rad/s) | | Magnitude of CMGs momentum error (Nms) | | Magnitude of total momentum error (Nms) | |
|---|---|---|---|---|---|---|---|---|
| | Average | Maximum | Average | Maximum | Average | Maximum | Average | Maximum |
| Tradition tracking | $0.4805\times10^{-3}$ | $0.6413\times10^{-3}$ | $0.3709\times10^{-7}$ | $0.4438\times10^{-7}$ | $3.6156\times10^{3}$ | $7.0373\times10^{3}$ | $3.6156\times10^{3}$ | $7.0379\times10^{3}$ |
| Robust tracking with LTAC | 0.0754 | 0.3267 | $0.1626\times10^{-5}$ | $0.7032\times10^{-5}$ | $1.7560\times10^{2}$ | $7.6171\times10^{2}$ | $1.3172\times10^{2}$ | $5.7146\times10^{2}$ |
| Robust tracking with RTAC | 0.0026 | 0.0627 | $0.0082\times10^{-5}$ | $0.1795\times10^{-5}$ | 1.4483 | 11.0783 | 1.3665 | 34.4581 |

### 4.2 Maneuver under environmental torque modeling error

The real environmental torque acting on the space station is different from that used in the trajectory planning. Consider a disturbance environmental torque, which is not included in the trajectory planning model and is given by

$$\boldsymbol{\tau}_{dis}(t) = \boldsymbol{v}_d \left( \frac{2(t - t_0)}{t_f - t_0} \right)^6 \left( \frac{2(t - t_0)}{t_f - t_0} - 2 \right)^6 \tag{33}$$

where $\boldsymbol{v}_d = [4, 4, 4]^T$ Nm. $\boldsymbol{v}_d$ is intentionally set large to investigate the performance of guidance strategies. The maneuver simulation is initialized with the exact nominal initial boundary conditions. Figures 7-9 compare the simulated attitude profiles, and Fig. 10 gives the magnitude of the CMGs momentum error. For the traditional tracking, the nominal attitude trajectory is well tracked, at the price of the CMGs momentum error. For the robust tracking with LATC, the terminal error of the CMGs momentum decreases greatly but it still brings obvious terminal attitude error. For the robust tracking with RATC, the performance is satisfactory; the principal rotation angle of terminal attitude error is 0.099 deg and the terminal CMGs momentum error is 65.7839 Nms.

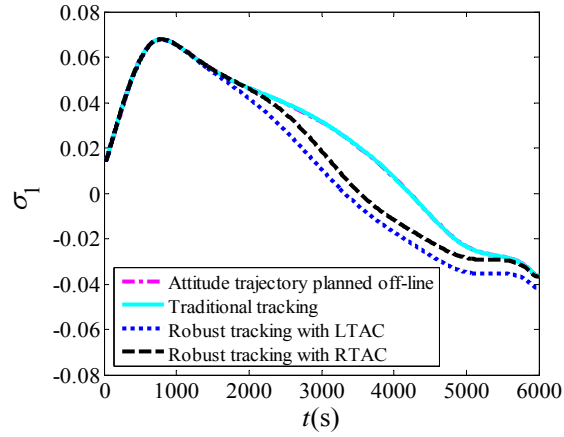

**Fig. 7. Simulation results for $\sigma_1$ under the disturbance torque**

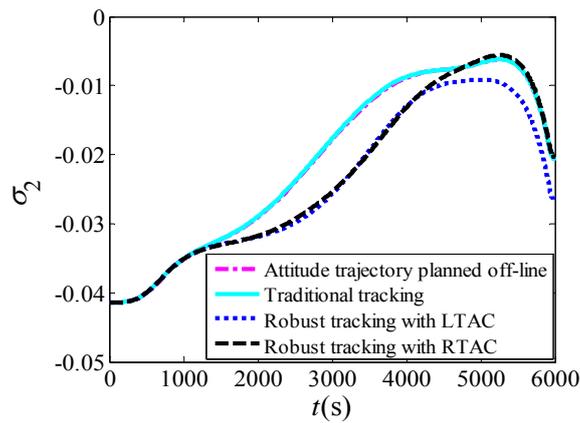

**Fig. 8. Simulation results for $\sigma_2$ under the disturbance torque**

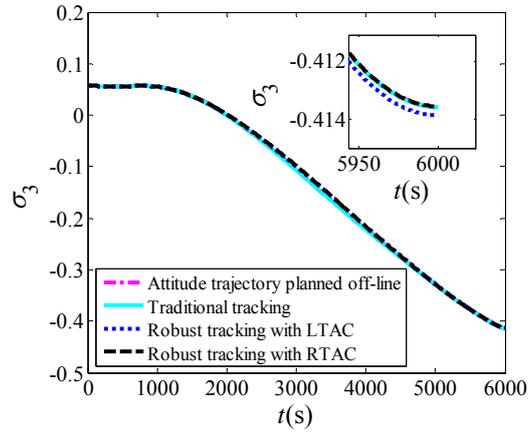

**Fig. 9. Simulation results for $\sigma_3$ under the disturbance torque**

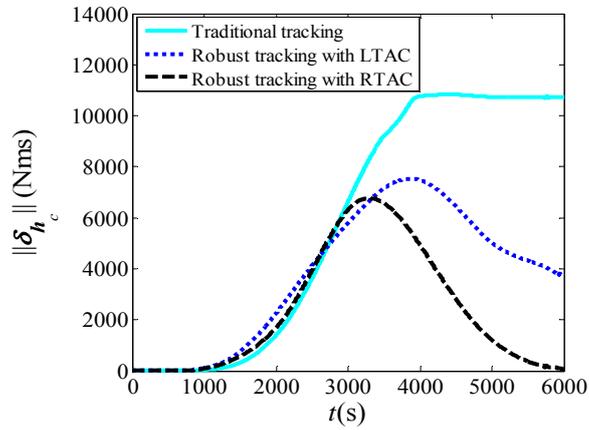

**Fig. 10. Simulation results of the CMGs momentum error magnitude under the disturbance torque**

Figure 11 plots the momentum magnitude of the CMGs in the maneuver. Under the disturbance environmental torque, the CMGs momentum saturation occurs using the traditional tracking, while this is avoided using the robust tracking guidance.

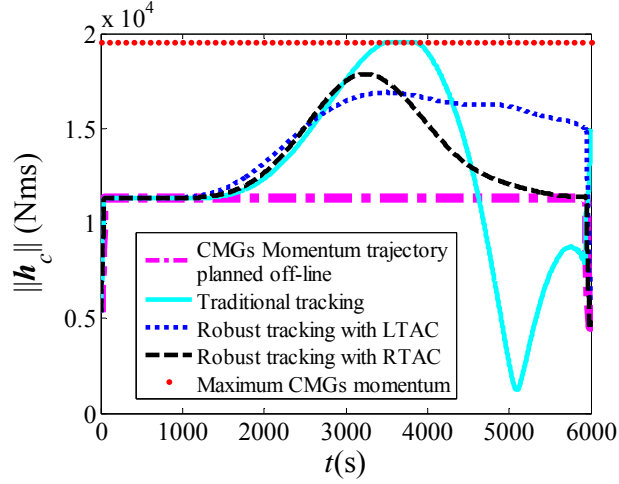

**Fig. 11. Simulation results of the CMGs momentum magnitude under the disturbance torque**

Monte-Carlo simulations with 100 samples were also performed. The disturbance torque given by Eq. (33) is considered with $v_d = [1.5, 1.5, 1.5]^T$ Nm. The set of initial state errors is same as that in Section 4.1. Table 4 presents the terminal state errors, and shows that the robust guidance strategy with RATC finishes the maneuver in high quality.

**Table 4. Terminal state errors of ZPM under disturbance torque and initial state errors**

| Tracking Strategy | Principal rotation angle of attitude error (deg) | | Magnitude of angular velocity error (rad/s) | | Magnitude of CMGs momentum error (Nms) | | Magnitude of total momentum error (Nms) | |
|---|---|---|---|---|---|---|---|---|
| | Average | Maximum | Average | Maximum | Average | Maximum | Average | Maximum |
| Tradition tracking | $0.4733 \times 10^{-3}$ | $0.5850 \times 10^{-3}$ | $0.3736 \times 10^{-7}$ | $0.4491 \times 10^{-7}$ | $0.5808 \times 10^{4}$ | $1.1110 \times 10^{4}$ | $0.5807 \times 10^{4}$ | $1.1109 \times 10^{4}$ |
| Robust tracking with LTAC | 0.6478 | 0.9310 | $0.1430 \times 10^{-4}$ | $0.2039 \times 10^{-4}$ | $1.5254 \times 10^{3}$ | $2.1862 \times 10^{3}$ | $1.1449 \times 10^{3}$ | $1.6401 \times 10^{3}$ |
| Robust tracking with RTAC | 0.0373 | 0.0410 | $0.1478 \times 10^{-5}$ | $0.1581 \times 10^{-5}$ | 25.4540 | 36.9666 | 22.0136 | 23.9109 |

### 4.3 Maneuver under inertia matrix uncertainty

In this subsection, the situation, that the actual inertia property of space station in the maneuver is different from the nominal used in the trajectory planning, is considered. It is presumed that the actual inertia matrix may be obtained on-line as realized in Paynter and Bishop (1997). In the simulation, the actual inertia matrix is $1.10\boldsymbol{J}$, where $\boldsymbol{J}$ is the nominal matrix given by Eq. (31). Simulated under the nominal initial states and environmental torque, the magnitude of the CMGs momentum is given in Fig. 12. The terminal CMGs momentum errors, for the traditional tracking, the robust tracking with LTAC, and the robust tracking with RTAC, are $1.2099\times10^3$, $1.2809\times10^2$, and $0.9617$ Nms, respectively. The terminal attitude error for the robust tracking with RTAC is also very small, only $0.4787\times10^{-4}$ deg in principal rotation angle.

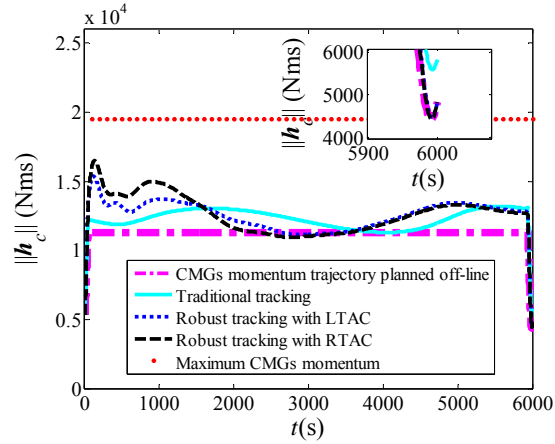

**Fig. 12. Simulation results of the CMGs momentum magnitude under the inertial uncertainty**

Monte-Carlo simulations with 100 samples were carried out. The principle moment of inertia falls into a uniform distribution ranging from 95% to 105% of the nominal value. The random initial state errors are considered; the principal rotation angle of attitude error is set as 3 deg, the magnitude of angular velocity error is 0.03 $n$, and the CMGs momentum error magnitude is 800 Nms. The disturbance torque is considered with $v_d = [1.5, 1.5, 1.5]^T$ Nm in Eq. (33). Table 5 presents the terminal state errors, and again shows the reliability of robust guidance strategy with RATC.

**Table 5. Terminal state errors of ZPM under inertia uncertainty, disturbance torque and initial state errors**

| Tracking Strategy | Principal rotation angle of attitude error (deg) | | Magnitude of angular velocity error (rad/s) | | Magnitude of CMGs momentum error (Nms) | | Magnitude of total momentum error (Nms) | |
|---|---|---|---|---|---|---|---|---|
| | Average | Maximum | Average | Maximum | Average | Maximum | Average | Maximum |
| Tradition tracking | 0.2159 | 21.5192 | $0.0052\times10^{-3}$ | $0.4460\times10^{-3}$ | $0.5468\times10^4$ | $1.9615\times10^4$ | $0.5654\times10^4$ | $3.5474\times10^4$ |
| Robust tracking with LTAC | 0.6669 | 1.3254 | $0.1484\times10^{-4}$ | $0.2725\times10^{-4}$ | $1.5680\times10^3$ | $3.3413\times10^3$ | $1.1803\times10^3$ | $2.4927\times10^3$ |
| Robust tracking with RTAC | 0.0741 | 0.8646 | $0.0204\times10^{-4}$ | $0.1339\times10^{-4}$ | 38.3786 | 383.0331 | 49.8116 | 650.9651 |

For all the maneuvers performed in the paper, the CMGs momentum saturation occurs sometimes for both the traditional tracking guidance and the robust tracking guidance. However, when using the robust tracking with RTAC, all maneuvers may be regarded successful because of the tolerable terminal error. This proves that once the error in the total angular momentum is effectively attenuated, the violation of path constraints will be eliminated.

## 5. Conclusion

A robust tracking guidance strategy for the Zero Propellant Maneuver (ZPM) is proposed. By guiding the total angular momentum of the space station system to the expected state, the space station is maneuvered to the target states through the reference attitude trajectory adjusted on-line, and the disturbance effects arising from the initial state errors, the environmental torque modeling error and the inertia uncertainty may be attenuated or even eliminated. The robust tracking guidance enhances the performance of ZPM, and is meaningful to achieve the goal that standard attitude trajectories may be stored on the space station in advance for regular maneuver missions.

Although the target states are well achieved for the ZPM guidance problem, it is recognized that in the work the path constraints originating from the Control Momentum Gyroscopes (CMGs) performance limit are not satisfactorily solved. They may be active during the maneuver. Avoiding the violation of the path constraints is significant for the safety of the ZPM, and it presents the challenge of addressing complex path constraints in the control realization. This needs to be investigated in the further studies. In spite of that, the proposed method effectively eliminates the error in the total angular momentum during the maneuver, which determines that path constraints regarding the CMGs will ultimately be satisfied as planned, and the momentum-optimal path selected is advantageous to reduce the occurrence of path constraint violation. The large amount of simulations proves the superiority to the traditional tracking. Especially, to the most of the authors' knowledge, this is a first attempt of regulating the reference trajectory in tracking without using a planning method. Certainly there will be a lot of details to address for the real flight, such as the computational architecture in software or even the hardware on board, yet undoubtedly the work presents a promising way to evolve the ZPM technique. In addition, for the regulation is a trivial case of tracking, the control strategy is also applicable to the momentum management of the space station.

## Acknowledgments

This research was supported by the National Natural Science Foundation of China (11272346) and the National Key Basic Research and Development Program (2013CB733100).